\documentclass[11pt,a4paper]{article}
\usepackage[hyperref]{emnlp2018}
\usepackage{times}
\usepackage{latexsym}
\usepackage{times}
\usepackage{xcolor}
\usepackage{soul}
\usepackage[utf8]{inputenc}
\usepackage[small]{caption}
\usepackage{amsmath}
\usepackage{booktabs}
\usepackage{verbatim}
\usepackage{latexsym}
\usepackage{url}
\usepackage{multirow}
\usepackage{listings}
\usepackage{color}
\usepackage{colortbl}
\usepackage{todonotes}
\usepackage{lscape}
\usepackage{rotating}
\usepackage{epstopdf}
\usepackage{hyperref}
\usepackage{amssymb} 
\usepackage{subcaption}
\usepackage{cleveref}
\usepackage{comment}
\usepackage{tabularx,ragged2e,booktabs}
\newcolumntype{L}{>{\RaggedRight\arraybackslash}X} 
\usepackage{pgfplots}
\usepackage{enumitem}
\usepackage{epstopdf}
\usepackage{CJKutf8}
\usepackage{graphicx}
\usepackage{mathtools}
\usepackage{flexisym}

\newcolumntype{P}[1]{>{\centering\arraybackslash}p{#1}}
\newcolumntype{L}[1]{>{\raggedleft\arraybackslash}p{#1}}
\newcolumntype{R}[1]{>{\raggedright\arraybackslash}p{#1}}

\graphicspath{{graphs/}{imgs/}}

\usepackage[ruled,vlined,linesnumbered]{algorithm2e}

\usetikzlibrary{shapes.geometric, arrows}

\definecolor{LightCyan}{rgb}{255,255,204}
\aclfinalcopy 

\title{Named Entity Recognition on Noisy Data using Images and Text\\(1-page abstract)}

\author{Diego Esteves \\
  SDA Research, University of Bonn, Germany
  \\
  {\tt esteves@cs.uni-bonn.de} \\}

\date{}

\begin{document}
\maketitle

\begin{figure}[ht]
\centering
\includegraphics[scale=0.35]{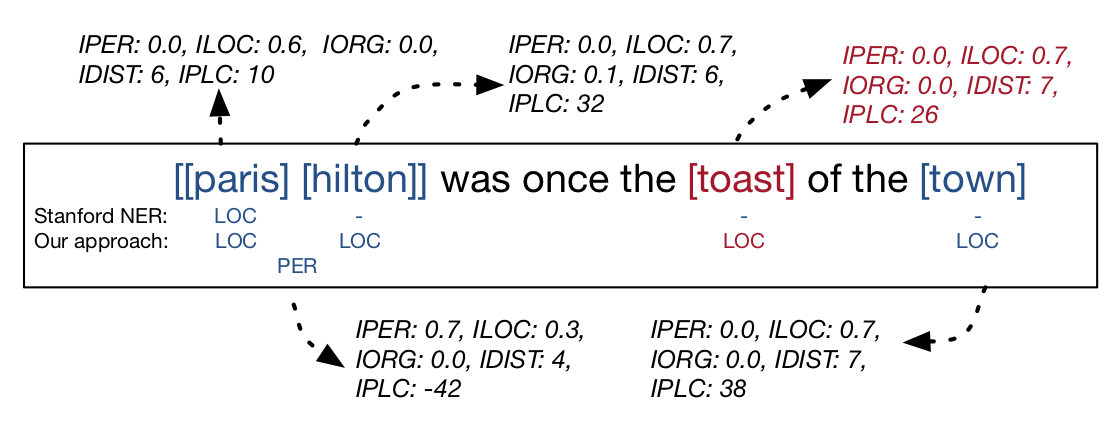}
\caption{Example of extracted features.}
\label{fig:example}
\end{figure}

\begin{figure}[ht]
\centering
\includegraphics[scale=0.25]{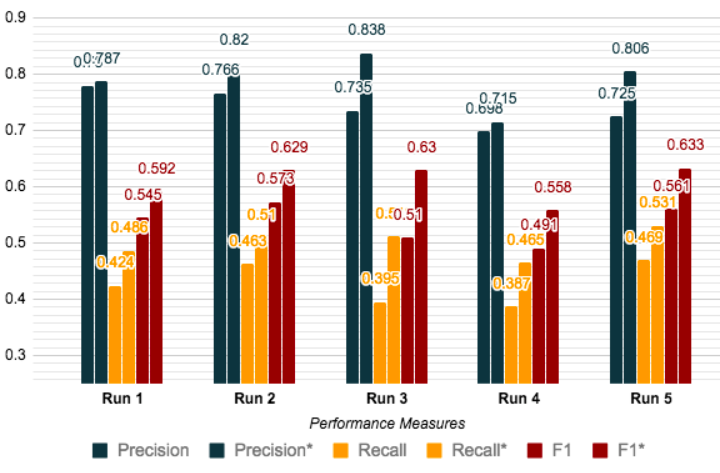}
\caption{Performance measures in Ritter dataset (3-MUC).}
\label{fig:performance}
\end{figure}

\textbf{Motivation} The importance of real-world knowledge (a.k.a. \textit{common sense}) for NLP was first discussed as early as 1960~\cite{bar1960present}
. However, up to now a substantial fraction of problems involving NLP could only be fully resolved if a rich understanding of the world is available
\textbf{Current Advances} Recently, deep neural network architectures have achieved one step further in NER on noisy data - successfully overcoming the dependency of \textit{gazetteers} and encoded rules~\cite{limsopatham2016bidirectional} - but are still far from performing as good as in formal language domains\footnote{NER over newswire often perform (F1) above $0.90~0.95$.}. For instance, SOTA NER have (AVG) F1-measure ranging from $0.30$ to $0.50$ depending on the number of classes and the dataset, which confirms the very challenging characteristic of the task in noisy data. In short, this occurs due to the lack of linguistic formalism. Findings of a recent and comprehensive qualitative study of this gap are presented by~\cite{AUGENSTEIN201761}. \textbf{Embedding world-knowledge} The main insight underlying the proposed work\footnote{Esteves et al. 2017. \textit{Named Entity Recognition in Twitter Using Images and Text}. Current Trends in Web Engineering - {ICWE} 2017 International Workshops.} is that we can enrich NER models by adding global features extracted from images and text in a word level perspective. To this end, we train a set of computer vision classifiers to recognize a set of pre-defined objects (each set associated to each named entity) as well as a set of text classification classifiers to label documents.
A sentence example of the produced features extracted by the computer vision module is shown in~\Cref{fig:example}. 
Results of the performance (5-fold) of a 3-classes CRF model in Ritter is shown in~\Cref{fig:performance} ([*] with features), confirming the potential of this approach. 
\textbf{Work in progress} The preliminary version of this framework uses TF-IDF based text classification and SIFT features for computer vision. These components are being extended with more robust algorithms such as Topic Modeling and CNNs. Furthermore, due to its comprehensiveness, we are re-training the model using a recently released corpus: The Broad Twitter Corpus
as well as extending the coverage of NE classes.

\bibliographystyle{acl_natbib_nourl}
\bibliography{emnlp2018}

\begin{thebibliography}{3}
\expandafter\ifx\csname natexlab\endcsname\relax\def\natexlab#1{#1}\fi

\bibitem[{Augenstein et~al.(2017)Augenstein, Derczynski, and
  Bontcheva}]{AUGENSTEIN201761}
Isabelle Augenstein, Leon Derczynski, and Kalina Bontcheva. 2017.
\newblock Generalisation in named entity recognition: A quantitative analysis.
\newblock \emph{Computer Speech \& Language}.

\bibitem[{Bar-Hillel(1960)}]{bar1960present}
Yehoshua Bar-Hillel. 1960.
\newblock The present status of automatic translation of languages.
\newblock In \emph{Advances in computers}, volume~1, pages 91--163. Elsevier.

\bibitem[{Limsopatham and Collier(2016)}]{limsopatham2016bidirectional}
Nut Limsopatham and Nigel Collier. 2016.
\newblock Bidirectional lstm for named entity recognition in twitter messages.
\newblock \emph{WNUT 2016}, page 145.

\end{thebibliography}

\end{document}